\documentstyle[twocolumn,aps,prb,epsf]{revtex}
\begin{document}

\title{PROTEIN FOLDING AND MODELS OF DYNAMICS ON THE LATTICE}

\author{Trinh Xuan Hoang and Marek Cieplak}

\address{Institute of Physics, Polish Academy of Sciences,
02-668 Warsaw, Poland}

\address{
\centering{
\medskip\em
{}~\\
\begin{minipage}{14cm}
We study folding in 16-monomer heteropolymers on the square lattice.
For a given sequence, thermodynamic properties and stability of the
native state are unique. However, the kinetics of folding depends on
the model of dynamics adopted for the time evolution of the system. We
consider three such models: Rouse-like dynamics with either single
monomer moves or with single and double monomer moves, and the
'slithering snake' dynamics. Usually, the snake dynamics has poorer
folding properties compared to the Rouse-like dynamics, but examples
of opposite behavior can also be found.  This behavior relates to
which conformations act as local energy minima when their stability is
checked against the moves of a particular dynamics.  A
characteristic temperature related to the combined probability, $P_L$,
to stay in the non-native minima during folding
coincides with the temperature of the
fastest folding.  Studies of $P_L$ yield an easy numerical way to
determine conditions of the optimal folding.
{}~\\
{}~
\end{minipage}
}}

\maketitle
\newpage

Proteins that are found in nature fold rapidly to their native states
when physiological conditions are restored \cite{Anfinsen,Dill}.
Random sequences of amino acids, on the other hand, may take forever to
fold \cite{Levinthal} or they may have a non-compact ground state.
Lattice models have provided insights into the
key problems of folding kinetics like the transition through the
compactification stage \cite{Chan} and existence and characterization of
the folding funnel \cite{rank,cell,Master,Coarse}.  For the
lattice models, the dynamics needs to be declared and there are various
ways to  define the single step moves for a given Hamiltonian. The
time evolution is then implemented by performing a Monte Carlo process
or by using the Master equation \cite{Master}.  One usually adopts the
Rouse-like dynamics \cite{Rouse,see} in which there are two kinds of
motions: single- and double-monomer moves. The single-monomer move
consists of end-flip and corner moves while the double-monomer move
consists of the crankshaft-like rotation. Typically, one declares a
certain proportion in which the two kinds of motions are attempted.
For instance, one attempts single monomer moves with probability 0.2
and the double monomer moves with probability 0.8 \cite{see}.  Chan
and Dill\cite{Chan} have also studied an expanded set of moves in
which a rotation of large segments of the polymer was also allowed.

The thermodynamic stability of a lattice heteropolymer in its native
state depends only on the energy spectrum, i.e. on the Hamiltonian,
but not on the dynamics itself.  The thermodynamic stability may be
characterized by the folding temperature, $T_f$, defined by the value
of temperature, $T$, at which the equilibrium probability to fold,
$P_0$, is $\frac{1}{2}$.  The kinetic propensity to fold may be
characterized by $T_{min}$ -- the temperature at which the folding
process is the fastest, or by the glass transition temperature, $T_g$,
below which the kinetics becomes so slow that folding is kinetically
unlikely \cite{Socci}.   The definition of $T_g$ relies on the cutoff
value of the characteristic folding time whereas $T_{min}$ is defined
uniquely and it seems preferable to use the latter. Below $T_{min}$,
an onset of the glassy effects takes place and the value of $T_{min}$
depends on the dynamics.  Here, we demonstrate that this dependence is
significant. In particular, we show that a sequence may not even fold
if the dynamics is not chosen adequately.  Furthermore, we demonstrate
that $T_{min}$ is related to the combined probability, $P_L$, for the
sequence to be in non-native local energy minima  before finding the
native state. Specifically, we show that $T_{min}$ coincides with the
temperature at which $P_L$ crosses $\frac{1}{2}$.  Notice that whether
a given conformation is a local energy minimum or not depends on the
set of the dynamics moves used because existence of stability against
these moves is what defines a minimum. 

We study several heteropolymers on the square lattice and we  consider
three models of the dynamics: 1) standard Rouse-like dynamics (RD)
with the single and double moves applied with the proportions
mentioned above, 2) single monomer dynamics (RD1), and 3)
the 'slithering snake' dynamics  introduced by Wall and Mandel (WMD --
for Wall-Mandel dynamics) \cite{Wall}. The latter model imitates
snake-like displacement that characterizes motion of polymers in dense
solutions and has been introduced as a numerical implementation of the
reptation model proposed by De Gennes \cite{Gennes}.  It is also
related to diffusion of a mobile chain through an ordered array of
immobile obstacles.  Briefly, the dynamics involves choosing randomly
one end of the chain and then attempting to advance it to a new
neighboring lattice site with the remaining chain following along the
previous contour.  On the square lattice, both head and tail can
attempt to move to three new destinations each. The motion is not
allowed if the end site would be occupied by the chain {\em after} the
whole slithering displacement was accomplished.  An example of the
snake move is represented in Figure 1a.  This kind of the dynamics is
known \cite{Ebert} to lead to the well defined $t^{1/4}$ law for the
mean square displacement of the central bead, and then to a Rouse-like
$t^{1/2}$ law, before the asymptotic diffusive behavior is reached.
Here, $t$ denotes time.  Our motivation to consider the snake dynamics
is mostly formal -- we would like to discuss a dynamics which is
clearly distinct from RD. It is conceivable, however, that there may
exist sufficiently dense environments in which reptation may turn out
to represent the protein motion better.

\begin{figure}
\epsfxsize=3in
\centerline{\epsffile{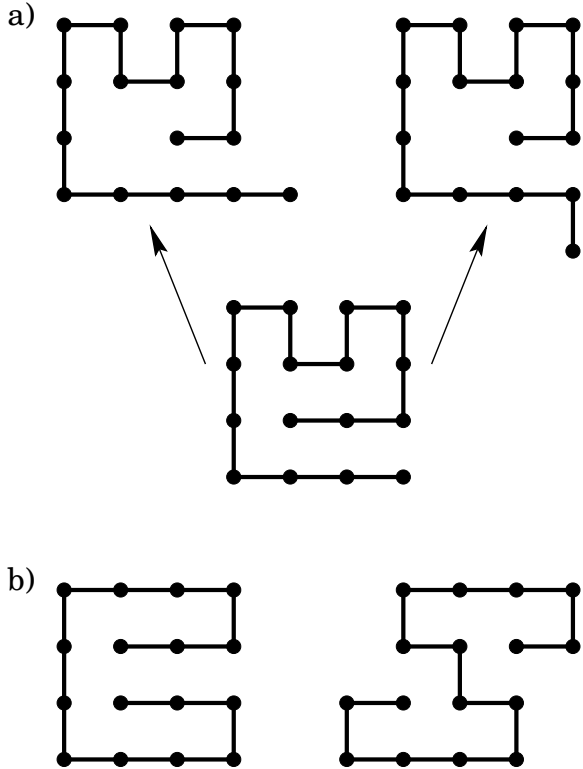}}
\vspace{10pt}
\caption{
a) An example of the snake move in WMD: starting from the conformation
at the bottom the chain can make a motion either to two adjacent
conformations on the top-left and the top-right of the figure.   The
end bead of the chain finds a new position on the lattice and all the
other beads `slither' forward along the previous contour by one
lattice constant. Note that the conformation at the bottom is also the
native conformation of sequence R. 
b) Non-ergodicity effects in 16-mer chain: the conformation on the
left is not accessible from any other
conformation for the move set present in RD, the same thing happens in
WMD to the conformation on the right. 
}
\end{figure}

Another issue is that of ergodicity. 
As pointed out by e.g. Chan and Dill \cite{Chan},
the move set of Rouse-like dynamics is not ergodic for
16-mer chains on the square lattice. We can see this by considering
the conformation shown on the left of Figure 1b.  This conformation
can never be reached by the RD moves but it can by the WMD moves. The
conformation on the right of Figure 1b shows a behavior which is just
the opposite.  For longer chains, non-ergodicity may become
significant.

We consider three 16-monomer sequences on the square lattice.
Two of them, R and DSKS', have couplings generated with 
the Gaussian probability distribution and their values 
are listed in Ref. 6.
Sequence R is constructed by the rank-ordering
technique that assigns the most strongly attractive couplings to the
native contacts in a target structure \cite{cell}. This sequence has
been found to be a good folder under the RD \cite{Coarse}. 
Sequence DSKS', first studied by Dinner et al. \cite{DSKS},  is a bad
folder within the same dynamics.   We demonstrate that, under WMD,
both sequences become bad folders.  We then consider an HP-sequence
\cite{Chan}, which we shall encode as HP2, since it has two 2 polar
and 14 hydrophobic beads, for which WMD provides better folding than
RD. This sequence has the structure H-H-H-H-P-H-H-H-H-H-H-P-H-H-H-H
and the corresponding native state is shown in Figure 2a.

\begin{figure}
\epsfxsize=3in
\centerline{\epsffile{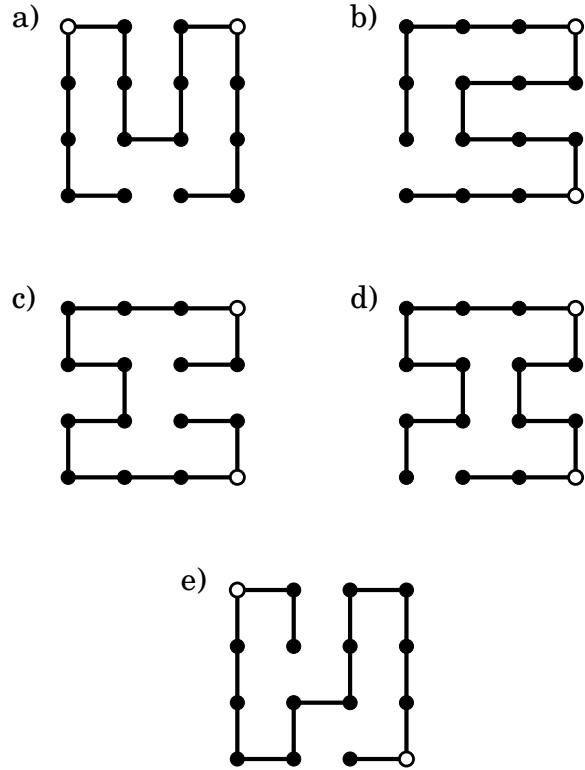}}
\vspace{10pt}
\caption{Native conformations of selected HP sequences with 2 polar
beads. The filled and open circles denote hydrophobic (H) and
polar (P) amino acids respectively.  2a) corresponds to sequence HP2.
}
\end{figure}

In lattice models, an energy of a sequence in a conformation is given by
\begin{equation}
E\; = \; \sum _{i<j} \; B_{i,j} \; \Delta (i-j) \;\;\;,
\end{equation}
where $\Delta(i-j)$ denotes presence of a contact between monomers $i$
and $j$, i.e. $\Delta(i-j)$ is 1 if indices $i$ and $j$ belong to
beads that are nearest neighbors on a lattice but are not neighbors
along the sequence.  Otherwise, $\Delta (i-j)$ is set equal to 0.
$B_{i,j}$ are the corresponding contact energies. Basically, in
Gaussian model, $B_{i,j}$'s have Gaussian values 
with a mean shifted by negative numbers
to provide compactness in the ground state. 
In the HP model
\cite{Dill} there are only three types of contacts and their energies
are $-1, 0, 0$ for the H-H, H-P and P-P pairs respectively.  The
values of $T_f$ are obtained by an exact enumeration of all
conformations and are equal to $1.15$, $0.195$, and $0.164$ for
sequences R, DSKS', and HP2 respectively.

Our Monte Carlo simulations have been done in a way that satisfies the
detailed balance conditions \cite{Master} and were devised along the
lines described in ref. \cite{Chan}. For polymers, satisfying these
conditions is non-trivial because each conformation has its own
number, $A$,  of allowed moves that the conformation can make. Thus
the propensities to make a move in a time unit vary from conformation
to conformation and the effective "activities" of the conformations
need to be matched. This can be accomplished by first determining the
maximum value of $A$, $A_{max}$.  For the 16-monomer chain on the
square lattice, $A_{max}$ is equal to 18, if the dynamics corresponds
to RD or RD1, and 6 in the case of WMD. We associate a single time
unit with the conformations in which $A$=$A_{max}$.   This means that
each allowed move is being attempted always with probability
$1/A_{max}$. For a conformation with $A$ allowed moves, probability to
attempt any move is then $A/A_{max}$ and probability not to do any
attempt is $1\;-\;A/A_{max}$.  The attempted moves are then accepted or
rejected as in the standard Metropolis procedure.  This description holds
for RD1 and WMD.  In the case of RD, probability to do a single
monomer move is additionally reduced by the factor of 0.2 and to do an
allowed double monomer move -- by 0.8.  The time used in Figures 3-5
is equal to the total number of the Monte Carlo attempts divided by
$A_{max}$.  This  scheme not only establishes the detailed balance
conditions \cite{Master} but it also uses less CPU compared to a
process in which moves are attempted with disregard to whether they
are allowed or not.

We have carried out Monte Carlo simulations to determine the
temperature dependence of the median folding time, $t_{fold}$, for the
three sequences and using the three models of the dynamics.  Figure 3
and 4 show the results for R and DSKS' and Figure 5 is for HP2.  For
each temperature the median folding time is determined based on 200
independent runs starting at random conformations (the cut off is set
at  value which is significantly above the lowest folding time). The
data points shown are averaged over 5 to 10 simulations, each 
corresponding to 200 trajectories. The figures show that $t_{fold}$ 
depends on the dynamics in a sensitive way but the temperature dependence
of $t_{fold}$ generally has a U-shape with a pronounced minimum (the
minimum becomes rather broad only for DSKS' with the RD1).  Sequence R
is a good folder within RD1 and especially RD but it becomes a bad
folder under WMD: $T_{min}$ is significantly above $T_f$.  DSKS' and
HP2, on the other hand, are both bad folders for all of the types of
dynamics considered here.  However, it is interesting to point out
that for HP2 the WMD dynamics yields a $T_{min}$ which is comparable
to that generated by RD and RD1 and the folding times themselves are
significantly reduced under WMD.  This situation, however, is
uncommon: in most cases that we studied, including other HP sequences,
except for those shown in Figures 2a - 2d, WMD tends to make the
folding poorer.  This is because a snake-like move usually breaks many
contacts.  Sequence HP2 is also uncommon in another respect: it folds
better under RD1 than under RD which suggests that for this sequence
the crankshaft moves are much less favorable than the single moves.

\begin{figure}
\epsfxsize=3.4in
\centerline{\epsffile{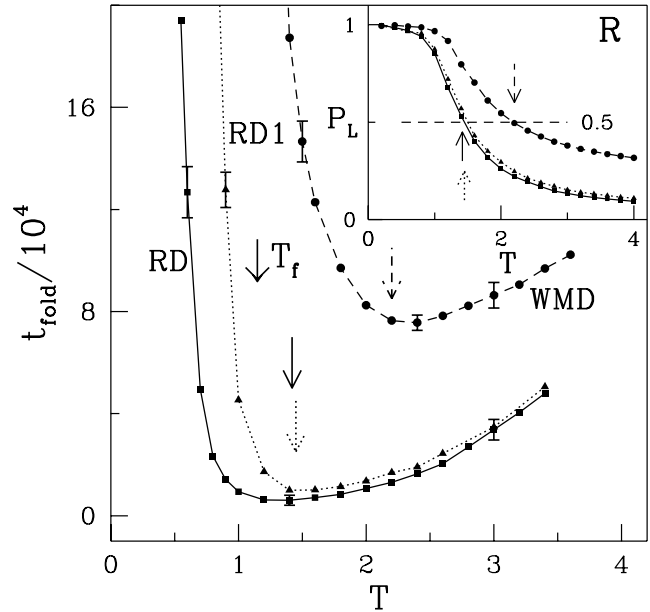}}
\vspace{1pt}
\caption{
The median folding time versus temperature for sequence R for the
three kinds of the dynamics:  RD, RD1 and WMD. The inset shows the
temperature dependence of $P_L$ . The arrow associated with $T_f$
indicates the folding temperature. The other arrows indicate
temperatures  at which $P_L$ crosses 0.5 for each type of the
dynamics. Note that they are very close to $T_{min}$.
}
\end{figure}

The geometry of the native conformation is also an important factor.
Generally an HP sequence folds better under WMD than under RD if it
has very few polar monomers, but this is not always the case.  Figure
2 shows the native states for several HP sequences with 2 monomers of
the P-type.  The first four of them are fast under WMD but the last
one (Figure 2e) is very slow.  We have checked that moving out of the
native state of Figure 2e by WMD involves a large energy barrier.

It is commonly accepted that folding is a motion that takes place in a
rugged energy landscape \cite{Dill}, which involves crossing many
energy barriers. The barriers arise due to the presence of local
energy minima (LM) in the system. The role of the minima can be
assessed by determining $P_L$ -- the probability to encounter LM's on
the way to folding.  This probability is defined as the fraction of
time spent in the LM's relative to the full folding time.  This
quantity depends on the dynamics explicitly: not only through the
definition of what conformation constitutes a minimum, which could be
analyzed by studying the energy spectrum, but also through the fact
that the associated weights are not necessarily Boltzmannian.  The
local minima themselves play a crucial role in some schemes to coarse
grain the description of the folding process \cite{Coarse}.

\begin{figure}
\epsfxsize=3.4in
\centerline{\epsffile{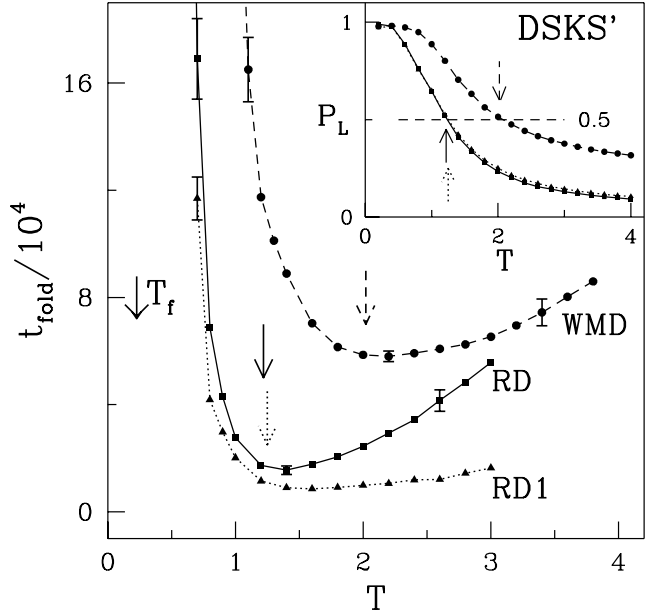}}
\vspace{1pt}
\caption{
Same as Figure 3 but for sequence DSKS'. 
}
\end{figure}
\begin{figure}
\epsfxsize=3.4in
\centerline{\epsffile{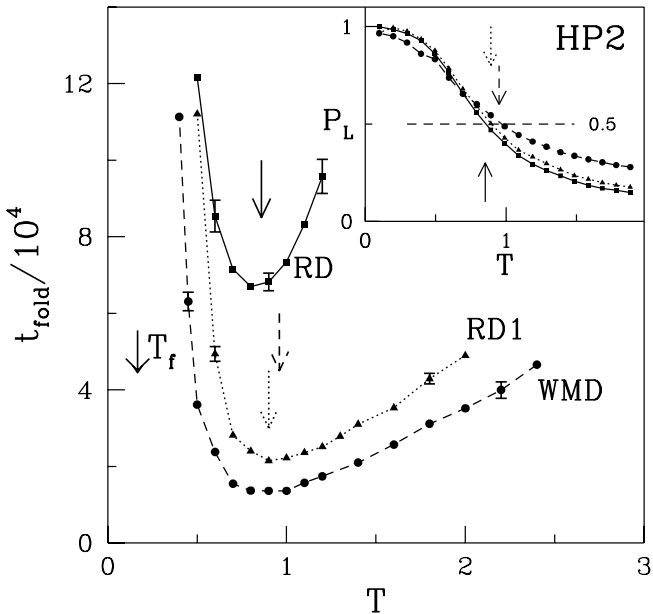}}
\vspace{5pt}
\caption{
Same as Figure 3  but for sequence HP2. 
}
\end{figure}

There are two kinds of conformations that are LM's: V-shaped,  if the
energy of the the conformation is lower than the energies of all
conformations that are immediately accessible from it, and U-shaped, if
one cannot reach states which are lower in  energy but some of the
allowed moves leave the energy unchanged.  For the 16-mer model there
are 802075 possible conformations and only a small fraction, $f$, of
these makes minima. With the RD dynamics, there are 9103 LM's for
sequence R out of which 2024 are U-shaped.  Each of the U-shaped
minima consists of several states thus the total number of states
involved in the U-shaped minima is 4893.  The total number of states
which are minima of whatever kind is then 11972 which makes about
$f$=1.5\% of the phase space.  The corresponding numbers for the other
sequences and other types of the dynamics are shown in Table I.
Checking whether a conformation  found on a Monte Carlo trajectory
is a local energy minimum or not
enhances the CPU by about 50\%. An incorporation of the
detailed balance conditions already involves an enumeration of the
possible moves but checking for the minima requires 
an additional determination
of the resulting energies.
Furthermore, checking wether the minimum is U-shaped requires
probing possible trajectories within a cutoff number of steps.

\vspace{20pt}
\begin{tabular}{c|c|c|c}
Sequence & RD & RD1 & WMD \\
\hline
R & 11\,972 (4\,893) & 16\,425 (8\,253) & 149\,443 \hfill\\
\hline
DSKS' & 12\,373 (5\,202)& 15\,851 (7\,846) & 150\,835 \hfill\\
\hline
HP2 & 12\,606 (5\,024) & 19\,142 (10\,093) & 103\,363 \hfill\\
\hline
\end{tabular}

\vspace{10pt}
\noindent
\begin{minipage}{3.4in}
\hspace{10pt}{\small TAB. 1. The total number of conformations that
are LM's for each of the sequences for the three kinds of dynamics. The
numbers in the brackets correspond to conformations which are
in the U-shaped LM's. In case of WMD there are no U-shaped LM's.}
\end{minipage}

\vspace{20pt}

For the WMD dynamics, the minima cut out an order of magnitude larger
portion of the phase space which alters the energy landscape
dramatically.  Notice also that when a chain makes a snake move the
set of new contacts usually
has no overlap with the preceding set of contacts.
In the Rouse-like dynamics, on the other hand, the conformations that
immediately connect to the native state have sets of contacts which
are overlapping to a large extent.

The small fraction of the phase space that corresponds to LM's freezes
the kinetics out at $T$=0. Thus at $T$=0 we get $P_L=1$ whereas at
high temperatures $P_L$ is of order $f$  -- as shown in Figure 6 for
sequence R and DSKS'.   There is then a crossover temperature $T_L$ at
which $P_L$ crosses $\frac{1}{2}$.  Notice that there is no such
crossover behavior for the quantity $P_L^{eq}$ which corresponds to
$P_L$ with the weights calculated at equilibrium -- through the
partition function.  The reason is that at $T$=0 it is only the native
state that has the occupation of 1.  Instead, $P_L^{eq}$ has a
maximum, which in the case of the sequence R, under RD coincides with
$T_{min}$.   In other cases, like for DSKS' as shown in Figure 6, the
peak in $P_L^{eq}$ is substantially displaced away from $T_{min}$ towards
lower temperatures.

\begin{figure}
\epsfxsize=3.4in
\centerline{\epsffile{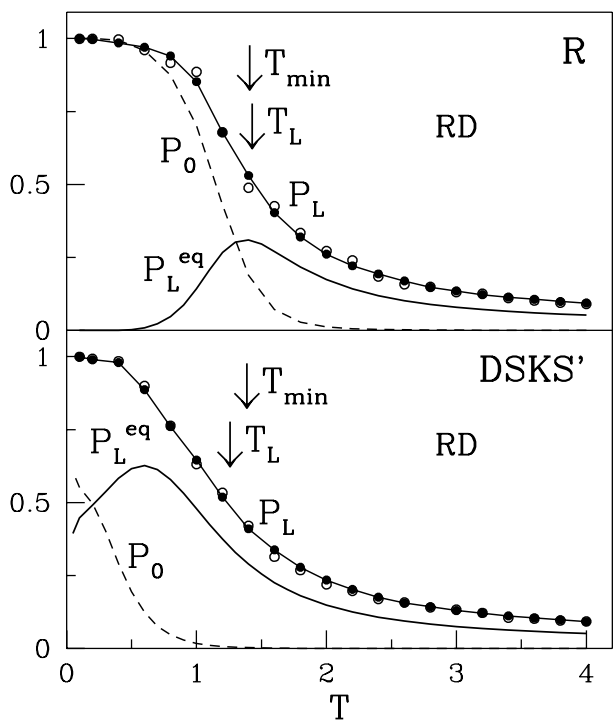}}
\vspace{5pt}
\caption{
Plots of $P_0$, $P_L$, and $P_L^{eq}$ vs. temperature for sequence R
(top) and DSKS' (bottom) under RD. $P_0$ and $P_L^{eq}$ are obtained
through the exact enumeration of states.  The data points for $P_L$
that are marked by the open circles are averaged over 5 MC
trajectories whereas those marked by black circles -- over 50
trajectories.
}
\end{figure}

The plots of $P_L$ vs. $T$ are shown in the insets of Figures 3,4, and
5. The data points shown are averaged over 50 trajectories.  The striking
observation is that $T_L$ appears to coincide with $T_{min}$ for any
sequence and for any model of the dynamics that we studied.  In other
words, folding turns out to be the most favorable at a temperature when
half of its time the sequence spends in the local minima during
folding. Thus $T_L$ is a measure of temperature below 
which kinetic trapping in the minima
becomes substantial. $T_L$ then conveys the same physics as
contained in $T_{min}$.  We have observed that $T_L$ is much easier to
calculate than $T_{min}$ because $P_L$, at any $T$, converges to a
well determined value quite fast: it becomes reliable already after
several runs -- as demonstrated in Figure 6.
For good folders,
the temperature at which $P_L^{eq}$ has a maximum
is expected to be somewhere around
$T_f$ because the maximum signifies an onset
of a substantial equilibrium occupation of the native state
and the folding funnel dominates the energy landscape.
For bad folders, on the other hand, we find that there is
essentially no relationship between the temperature
of the maximum and $T_f$ because the non-native minima
are blocking formation of the native funnel and the
position of the maximum is dominated by the nature of the
dynamics. 
The relationship between $T_L$ and $T_{min}$ is well defined
both for bad and good folders because the definitions 
of the two temperatures are anchored to the dynamics.

In summary, we have shown that the kinetics of folding strongly
depends on the details of the dynamics.  We have also provided a
simplified method to determine the temperature of the fastest folding.
The method is based on monitoring 
the combined occupation of the non-native local energy
minima. We have also indicated that, for good folders, 
the folding temperature can be estimated
by studying equilibrium occupation of the minima.
Thus the essential characteristics of 
well folding sequences can be obtained
by focusing on the energy minima instead of on the native state itself.
This may offer approximate ways to study longer sequences.

We thank for discussions with M. S. Li. This research has been
supported by KBN (Grant number 2P03B-025-13).


\end{document}